\documentclass{ametsocV6.2}

\title{Value of Adding AI Weather Model Ensembles to a Multi-model Probabilistic Blend}

\authors{Harris Sop Nkuiate,\aff{a}\correspondingauthor{Harris Sop Nkuiate, hn322@exeter.ac.uk} 
Gavin Evans,\aff{b} 
Christopher A. T. Ferro,\aff{a} 
Frank Kwasniok,\aff{a}
Caroline Bain,\aff{b} 
and Stefan Siegert,\aff{a} 
}

\affiliation{\aff{a}{Department of Mathematics and Statistics, University of Exeter, UK}\\
\aff{b}{Met Office, Exeter, UK}
}

\abstract{Many weather forecasting organizations use a multi-model blend of ensemble members from different prediction systems to obtain a more accurate forecast. Recent studies have shown that blending deterministic AI-based  models alongside physics-based models improves the overall forecast skill. However, the incorporation of an ensemble AI-based model into a multi-model blend is yet to be investigated. This study explores the value of blending the ensemble version of the Artificial Intelligence Forecasting System (AIFS-CRPS) with the Met Office's convection permitting physics-based ensemble, MOGREPS-UK. We show that blending site temperature (and wind speed) forecasts over the United Kingdom from both ensemble physics-based and ensemble AI-based models leads to an increase in overall forecast skill spatially, across lead times and across exceedance thresholds, especially when the models have different error characteristics. 
By decomposing the continuous ranked probability score of the blend into component means and diversity terms, we show that a more accurate blend can be achieved by assigning more weight to more skillful models and using models with complementary errors. For example, we show that a blend of MOGREPS-UK with either AIFS-CRPS or the physics-based Integrated Forecasting System (IFS) is better than a blend of IFS and AIFS-CRPS which are not very diverse. This has implications for the importance of weather model diversity and may lend weight to different model development approaches being pursued in tandem. We also demonstrate that calibrating AIFS-CRPS forecasts yields an increase in forecast skill comparable to calibrated ensemble physics-based forecasts, and calibrating the source forecasts before blending and the blended forecast after blending improves the overall forecast accuracy.} 
\begin{document}

\maketitle

%
%
%
%
%

%
\section{Introduction}

Recent advances in Machine Learning (ML) and Artificial Intelligence (AI) have led to the development of data-driven weather models whose forecast skills are comparable to those of physics-based models \citep{ben2024rise, price2025probabilistic}. AI-based weather models focus on learning non-linear patterns in historical observations, while physics-based models use numerical integration schemes to solve equations that govern the weather \citep{ben2024rise}. 

Although AI-based models require a large amount of data for training, they are computationally less expensive to run once trained and can produce larger ensembles than physics-based models with less computational cost \citep{de2023machine}. Examples of some well known deterministic AI-based models include Pangu-Weather \citep{bi2023accurate}, GraphCast \citep{lam2023learning}, FourCastNet \citep{pathak2022fourcastnet}, and Artificial Intelligence Forecasting System (AIFS) \citep{lang2024aifs}. 

\cite{lang2024aifs} noted that deterministic AI models tend to produce forecasts with smooth fields, thereby reducing the accuracy of these models. To address this challenge, \cite{lang2026aifs} proposed an ensemble version of AIFS called AIFS-CRPS, which uses the continuous ranked probability score (CRPS) as a loss function to account for uncertainties. Other well known probabilistic AI models include GenCast \citep{price2025probabilistic} and NeuralGCM \citep{kochkov2024neural}. Also, \citet{bulte2026uncertainty} proposed the use of uncertainty quantification (UQ) methods to generate probabilistic forecasts from deterministic AI models and \citet{gneiting2026probabilistic} contend for the use of potential CRPS to fairly compare these forecasts with their physics-based counterparts.

It is well known that ensemble forecasts from physics-based models are biased and under-dispersed and require the application of post-processing methods to address these issues \citep{gneiting2005calibrated}. However, there are limited studies on the application of post-processing to AI-based models. For example, \cite{bremnes2024evaluation} applied post-processing to Pangu-Weather to calibrate temperature and wind speeds using a neural network. \citet{trotta2025statistical} used reliability calibration \citep{flowerdew2014calibrating}, a non-parametric post-processing method which focuses on improving the forecast reliability while leaving its resolution unchanged, to correct for systematic errors in AIFS. \citet{hua2026improving} used a decoder only transformer (DOT) and a dense neural network (DNN) to post-process severe weather event predictions from Pangu-Weather and found that both methods lead to improved forecast performance, but DOT outperforms DNN. \citet{kocsis2026ai} applied ensemble model output statistics (EMOS) and quantile regression (QR) to AIFS-CRPS wind speed forecasts and found that both post-processing methods improve forecast skill and EMOS outperforms QR, especially at short lead times.

Multi-model blending is a technique used by operational weather centers to improve forecast skill by seamlessly combining different models based on their lead times. In the spirit of \citet{shipway2026blending}, this type of forecast blending falls under augmented machine learning, where the blending occurs outside the prediction system. \citet{rust2023improving} showed that a blend of the Met Office's physics-based regional and global models leads to an overall improvement in forecast performance. \citet{neal2024seamless} blended five global ensemble models using the Met Office's Decider post-processing system and showed that the multi-model blend outperforms the best performing individual model. Also, using the Brier score decomposition \citep{murphy1973new}, they demonstrated that blending improves both the reliability and resolution of the forecast. \citet{keller2021seamless} used the EMOS approach to blend temperature forecasts and found that the blend outperforms either of the single-model EMOS versions. 

\citet{trotta2024improving} used a simple neural network to blend forecasts from physics-based models and showed that non-linear weighting leads to more accurate blended forecasts compared to linear weighting. \citet{chakraborty2025mowe} proposed a mixture of experts approach in which different deterministic AI models are optimally blended using a Vision Transformer-based gating network. They obtained an RMSE of the blend that is up to 10$\%$ lower than the RMSE of the best performing model on a 2-day forecast horizon. \citet{trotta2025statistical} showed that adding deterministic AIFS to a multi-model blend improves forecast skill, even in cases where AIFS is less skillful than their physics-based model.

To the best of our knowledge, no analysis has been done to study the value of adding an ensemble AI-based forecast to the blend and the impact of post-processing before and after such blending. Therefore, the aim of this paper is to address these two questions using the Integrated Model Post-processing and Verification (IMPROVER) \citep{roberts2023improver} post-processing system and screen temperature forecasts. IMPROVER is open-source software and is the Met Office operational post-processing system, so the findings are highly relevant for future forecast design in the UK, but due to its use in other organizations worldwide, there may also be decision implications internationally. The remainder of this paper is structured as follows. Section 2 describes the different forecasts, including any preprocessing methods implemented, and their corresponding observations. Section 3 gives an overview of the IMPROVER processing and defines the different metrics used to assess the forecast performance. The results of the performance of both the individual forecasts and the blend will be presented in Section 4, while discussions and conclusions will be given in Section 5.

\section{Data}

The forecasts used in this study comprise site screen temperatures (and 10-m wind speeds in the Appendix) from 130 sites that span from 1 October 2024 to 31 January 2025 (123 days) at 24 hourly lead times and have validity times of 00Z. Data from 1 to 31 October 2024 is used for training and the remaining data (November to January) are used for evaluation. The physics-based models used include the Met Office's convection permitting ensemble, Met Office Global and Regional Ensemble Prediction System over the UK (MOGREPS-UK) \citep{hagelin2017met, porson2020recent}, the extended version (MOGREPS-G) \citep{porson2020recent}, and the European center for Medium-Range Weather Forecasts (ECMWF)'s Integrated Forecasting System (IFS) \citep{buizza2007new}. The AI-based forecasts are from the ECMWF's ensemble AI-based model, AIFS-CRPS \citep{lang2026aifs}. Both IFS and AIFS-CRPS forecasts are comprised of 50 ensemble members each, and the corresponding site observations are from synoptic observation reports from 112 World Meteorological Organization (WMO) synoptic observation stations around the UK. 

MOGREPS-UK forecasts are pre-processed onto a Lambert Azimuthal Equal Area projection with a horizontal resolution of 2 km, which is the grid used for operational post-processing at the Met Office. MOGREPS-G forecasts are pre-processed onto an equirectangular latitude-longitude projection with a horizontal resolution at midlatitudes of approximately 20km. The IFS and AIFS-CRPS forecasts are extracted from ECMWF’s Meteorological Archival and Retrieval System (MARS) and both pre-processed, as part of the extraction from MARS, to a latitude-longitude grid of $0.25^\circ$ resolution. The coarse resolution forecasts, MOGREPS-G, IFS and AIFS, are then regridded onto the same 2km UK grid used for MOGREPS-UK using land-sea aware, nearest neighbour interpolation. 

Site forecasts are created by selecting the nearest land grid point from the gridded forecasts. A lapse rate correction is applied to all forecasts to account for the difference in grid point average and site altitude. For MOGREPS-UK, where the NWP natively has a high horizontal resolution (2.2km, \citet{porson2020recent}), a local lapse rate is constructed using the screen temperatures and altitudes from the grid points surrounding the site of interest \citep{sheridan2010simple}, where the lapse rate is constrained to be between the dry adiabatic lapse rate (-0.01 Km$^{-1}$) and -3 times the Dry Adiabatic Lapse rate \citep{roberts2023improver}. For MOGREPS-G, IFS and AIFS, the nearest land grid point is selected with a fixed lapse rate used equal to the Environmental Lapse Rate (-0.0065 Km$^{-1}$).
\section{Methods}
\subsection{IMPROVER Processing}
MOGREPS-UK consists of 6 consecutive cycles (00Z, 23Z, 22Z, 21Z, 20Z, and 19Z) with each cycle having 3 ensemble members, while MOGREPS-G comprises 2 consecutive cycles (00Z and 18Z) with each having 18 ensemble members. These cycles are time-lagged (or termed cycle blending) to produce a total of 18 members for MOGREPS-UK  with lead times of T+24 to T+120, and 36 members for MOGREPS-G, with lead times of T+24 to T+198 both at 24-hour intervals \citep{porson2020recent}. Each cycle is given equal weight when blending consecutive cycles in order to reduce forecast oscillations \citep{griffiths2019flip}.

It is a common practice to eliminate systematic errors in forecasts before incorporating them in a multi-model blend by post-processing the individual forecasts separately at each site and lead time. We therefore implement the ensemble model output statistic (EMOS) method proposed by \citet{gneiting2005calibrated} to each site to correct these errors, since it is simple, robust, and computationally cheap. EMOS assumes a parametric distribution for each site and linearly relates the distribution's parameters to the ensemble statistics (mean and standard deviation). The model parameters are then estimated by minimizing the CRPS using the Nelder-Mead algorithm \citep{nelder1965simplex} for 200 iterations to limit the computational expense.

The predictive distribution of temperature, $y$, conditional on the ensemble members, $X_1,\ldots, X_m$, for EMOS is assumed to be a normal distribution
\begin{equation*}
    y|X_1,\ldots,X_m \sim \mathcal{N}(\mu, \sigma^2)
\end{equation*}
and the distribution parameters are linked to the ensemble mean, $\bar{X}$, and the standard deviation, $s$, through the functions
\begin{equation*}
\mu=a_0+a_1\bar{X}~~~\text{and}~~~\sigma^2=b_0+b_1s^2. 
\end{equation*}
The parameters of the model $a_0, a_1, b_0$ and $b_1$ are estimated using the training data, and $b_0$ and $b_1$ are constrained to be positive to ensure positive variance.

EMOS is then applied separately to MOGREPS-G and AIFS-CRPS to produce calibrated ensembles (sampled from the predictive distribution of EMOS), which are then used to investigate the benefits of calibrating forecast sources before blending.

IMPROVER produces probabilistic forecasts from ensembles by predicting the probabilities of exceedance for a given set of thresholds \citep{roberts2023improver}. That is, for any given threshold and ensemble forecast, it computes the proportion of ensemble members that exceeds that threshold. In this study, we choose thresholds ranging from $-10^\circ$C to $30^\circ$C with a step size of $1^\circ$C, making a total of 41 thresholds. Such a large number of thresholds is chosen so that the probability distribution can be modelled realistically. These exceedance probabilities for each forecast are then combined by taking the weighted average to form a multi-model blended probabilistic forecast.

Operationally, IMPROVER uses a linear weighting configuration that assigns weights based on the relative forecast skill of each model at different lead times. In this paper, we use a constant weighting configuration ($80/20$) with $80\%$ for the most skillful model and $20\%$ for the less skillful, since our goal is to investigate whether there is any value in adding an AI-based forecast to the blend. The skill of each model at a given lead time is decided by looking at their mean CRPS in the evaluation period.

Note that this weighting configuration can lead to jumps in forecast skill if the less skillful forecast suddenly becomes more skillful \citep{dabernig2024simplest}. This problem can be addressed in practice by using a gradient weighting configuration, where the weights are made to slowly increase or decrease across lead times to avoid temporal incoherence \citep{roberts2023improver}. 

Before the impact of calibration after blending is investigated, the calibrated blended probabilities are first converted to ensembles. We achieve this by interpolating between the probabilities specified using multiple thresholds to generate percentiles that are then used as ensemble members for post-processing \citep{flowerdew2014calibrating}.
EMOS is then applied to the derived ensembles to produce calibrated ensembles, which are then thresholded to obtain calibrated exceedance probabilities.

The deterministic forecasts (expected values) of the different models are derived from the probabilistic outputs by linearly interpolating these probabilities and integrating the resulting distribution. We now have exceedance probabilities and expected values for the raw and calibrated forecasts (MOGREPS-G and AIFS-CRPS) for the individual models and their blends, and we can assess their predictive performance using different metrics.

\subsection{Verification metrics}
\label{metrics}
Since IMPROVER produces probabilistic outputs (an approximation to a cumulative distribution function) for both individual forecasts and the blend, we will review some metrics used to evaluate the performance of such forecasts. The predictive performance of a probabilistic forecast can be assessed using a proper scoring rule. If a forecaster issues a distribution, $F$, and $y$ materialises, then their penalty is $S(F, y)$. We will consider scores to be negatively oriented in this work, that is, lower scores indicate higher reward, hence better forecast. There is a consensus in the forecasting community to use only scoring rules that are proper, as they ensure that the forecaster’s expected score is optimised only when they issue their true belief, thus promoting honesty \citep{gneiting2007strictly}. Score estimates from small ensembles are usually biased, and this can affect the predictive performance of a forecast. To overcome this problem, \citet{ferro2014fair} contends for the use of fair scores. However, we do not account for the difference in ensemble size when assessing the performance of the different forecasts in this study.

Suppose that we have a sequence of $N$ forecasts and their corresponding observations $\{(p_1,y_1),\ldots,(p_N,y_N) \}$, consisting of event probabilities $p_i\in[0,1]$ and the corresponding binary outcomes $y_i\in\{0,1\}$. The empirical average Brier score \citep{glenn1950verification} is defined by
\begin{displaymath}
    \overline{BS}=\frac{1}{N}\sum_{i=1}^{N}(p_i-y_i)^2.
\end{displaymath}
The Brier score measures the average square difference between the forecasts and their corresponding observations, and a smaller score is better. If we consider the case where the forecast can only take $K$ distinct values, $p_i\in\{f_1,\ldots,f_K\}$, then the average Brier score can be decomposed into three components using Murphy's decomposition \citep{murphy1973new}.

Let $n_k$ denote the number of times the forecast $f_k$ was issued and $o_k$ the total number of events that occurred when forecast $f_k$ was issued. The average event frequency given the forecasts $f_k$ is the ratio $o_k/n_k$ and the climatological event frequency, denoted by $\bar{o}$, is given by $(1/N)\sum_{i=1}^{N}y_i$. Murphy's decomposition of the average Brier score into reliability (REL), resolution (RES), and uncertainty (UNC) is given as
\begin{displaymath}
    \overline{\mathrm{BS}}=\mathrm{REL}-\mathrm{RES}+\mathrm{UNC}
\end{displaymath}
where
\begin{displaymath}
 \mathrm{REL}=\frac{1}{N}\sum_{k=1}^{K}n_k\left(\frac{o_k}{n_k}-f_k\right)^2,   
\end{displaymath}
\begin{displaymath}
 \mathrm{RES}=\frac{1}{N}\sum_{k=1}^{K}n_k\left(\frac{o_k}{n_k}-\bar{o}\right)^2,   
\end{displaymath}
and 
\begin{displaymath}
    \mathrm{UNC}=\bar{o}(1-\bar{o}).
\end{displaymath}
A forecast is calibrated if the predictive probabilities are close to their conditional event probabilities; thus, low values of $\mathrm{REL}$ indicate good performance in terms of calibration. This therefore suggests that one can construct a perfectly calibrated sequence of forecasts by issuing the climatology in all instances. However, such a sequence of forecasts would not be informative and would obtain $\mathrm{RES}=0$, which will negatively impact the score, since it cannot discriminate different scenarios better than climatology. 

The reliability term $\mathrm{REL}$ measures miscalibration as the difference in the mean score of the original forecasts to the calibrated forecasts. The resolution term $\mathrm{RES}$ on the other hand assesses the discrimination ability as the difference in the mean score of the calibrated forecasts to the reference forecast, for example the climatological forecast \citep{siegert2017simplifying}. This term therefore measures how much the distribution of observations changes when we condition on different forecasts. $\mathrm{UNC}$ measures uncertainty as the mean score with respect to the reference forecast. 

In the case where the forecast is a cumulative distribution function over the range of possible outcomes, the continuous ranked probability score (CRPS) \citep{matheson1976scoring, gneiting2007strictly} is the integral of the squared distance between the forecast distribution and the observation defined as follows:
\begin{displaymath}
  \mathrm{CRPS}(F,y)=\int_{-\infty}^{+\infty}[F(t)-\mathbbm{1}\{ t \geq y\}]^2 \,dt\\ \nonumber 
\end{displaymath}
where $F$ is the predictive cumulative distribution function and  $\mathbbm{1}$ is the indicator function which takes the value one when the observation $y$ does not exceed a given threshold $t$ ($t\geq y$) and zero otherwise. 

The improvement of the blended forecast relative to a reference forecast can be measured using skill scores \citep{murphy1973hedging}. Let $\mathrm{\overline{CRPS}}$ and $\mathrm{\overline{CRPS}_{ref}}$ denote the mean CRPS of the forecast being evaluated and the reference forecast, respectively. The continuous ranked probability skill score (CRPSS) is defined as
\begin{displaymath}
    \mathrm{CRPSS}=1-\frac{\mathrm{\overline{CRPS}}}{\mathrm{\overline{CRPS}_{ref}}}.
\end{displaymath}
Positive values indicate an improvement of the forecast over the reference forecast, while negative values indicate worse predictive performance. A skill score equal to zero suggests that there is no difference between the forecast under evaluation and the reference forecast.
\section{Results}
\subsection{Does adding AI-based forecasts into the blend increase the overall forecast performance?}
\begin{figure}[h]
\centerline{
 \includegraphics[width=33pc]{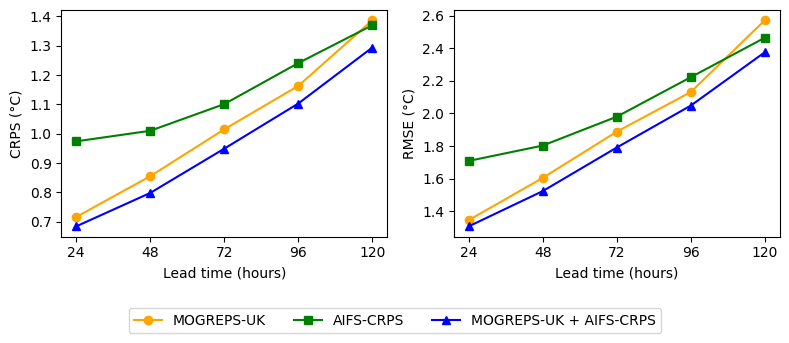}}
	\caption{CRPS and RMSE by forecast lead time for individual models and the blend of both averaged over all days and sites. Lower is better.}\label{fig1} 
\end{figure}
We begin our analysis by investigating whether there is value in blending AIFS-CRPS ensemble members with MOGREPS-UK ensemble members for 2m temperature over the UK. The mean CRPS and the root mean squared error (RMSE) of the expected value are calculated by averaging over all days and sites in the verification data, as shown in Figure \ref{fig1}. The figure shows that MOGREPS-UK outperforms AIFS-CRPS at all lead times for UK locations, except at day five, in terms of both deterministic and probabilistic performance. This means that AIFS-CRPS is becoming better than MOGREPS-UK at longer lead times. However, the blend performs better than the best performing individual model at all lead times. 

\begin{figure}[h]
\centerline{
 \includegraphics[width=39pc]{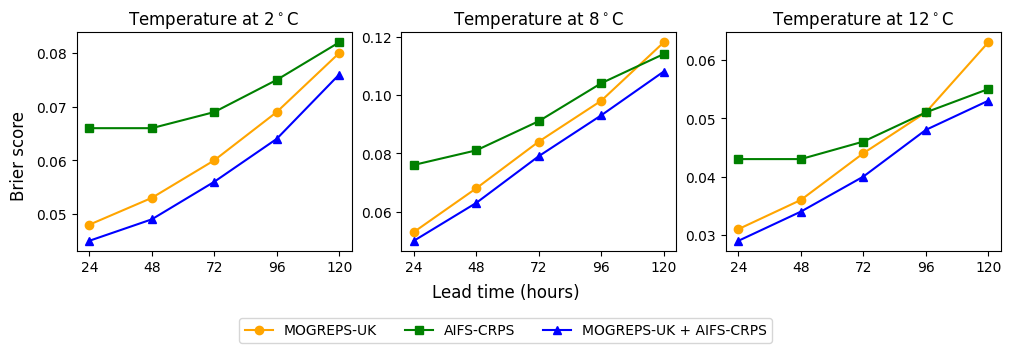}}
	\caption{Brier score at chosen thresholds against forecast lead time for individual models and the blend of both averaged over all days and sites. }\label{fig2} 
\end{figure}

Figure \ref{fig2} shows the Brier score at selected thresholds, roughly corresponding to the 10th, 50th, and 90th percentiles of the distribution of the observations pooled over all days and sites. At the lower threshold, MOGREPS-UK outperforms AIFS-CRPS at all lead times, while the blend has the best skill across all models at all lead times. Similarly, for mid- and higher thresholds, MOGREPS-UK outperforms AIFS-CRPS at all lead times, except at day five. Also, at these thresholds, the blend outperforms the best performing individual model on all lead times. 

Assessing the performance of both the overall and specific thresholds reveals that blending improves the accuracy of the forecasts whenever the forecasts are weighted according to their skill, with more weight given to the more skillful forecast, as in this study. Similar results are found when considering wind speed forecasts for the evaluation period, as shown in Figure \ref{appen1} in the appendix. Since the evaluation period covers mainly winter months, we investigate the behavior of the blend in the summer month of July 2025, as depicted in Figure \ref{appen2} (now again for temperature) in the appendix. This figure shows that the blend still outperforms the best performing individual model.

Our findings are consistent with those of \citet{trotta2025statistical}, where the blend comprises the deterministic AIFS and an ensemble physics-based model. Note that in \citet{trotta2025statistical}, fuzzy thresholding, where the predicted value is mapped to a
number between 0 and 1 depending on how far it is
from the threshold, is used to derive probabilistic forecasts from deterministic AIFS. This method is not as straightforward as using ensemble AIFS to obtain the probabilities of exceedance. 

\subsection{How does blending affect the forecast attributes?}
\begin{figure}[h]
\centerline{
 \includegraphics[width=39pc]{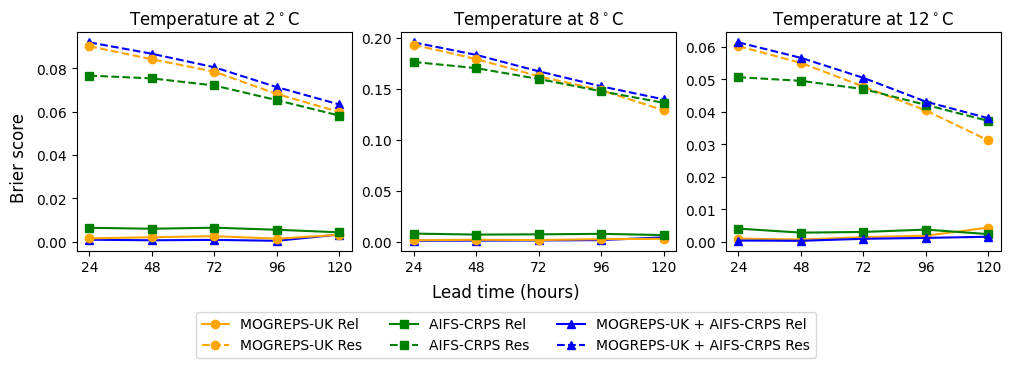}}
	\caption{The reliability (solid, lower is better) and resolution (dotted, higher is better) components of the mean Brier score averaged over all sites at chosen thresholds for individual models and the blend of both.}\label{fig3} 
\end{figure}

To understand the impact of blending on forecast properties in the blend, we use Murphy's decomposition of the mean Brier score described in Section 3\ref{metrics}. A smaller reliability term is better, while a larger resolution term is better. Figure \ref{fig3} shows that the resolution generally reduces with lead time and MOGREPS-UK has better resolution than AIFS-CRPS, especially at shorter lead times. 
However, the resolution of AIFS-CRPS surpasses that of MOGREPS-UK at day five for mid thresholds and days four and five for higher thresholds. The resolution of the blend outperforms the best performing model at all lead times and thresholds.

On the other hand, the reliability term does not show any appreciable trend across lead times, and MOGREPS-UK outperforms AIFS-CRPS at all lead times, except at day five. 
The reliability of the blend is similar to that of the better model in all cases. These results show that blending improves the reliability and resolution of the forecast, as observed in \citet{neal2024seamless}.

\subsection{Spatiotemporal performance of the blend}
\begin{figure}[h]
\centerline{
 \includegraphics[width=39pc]{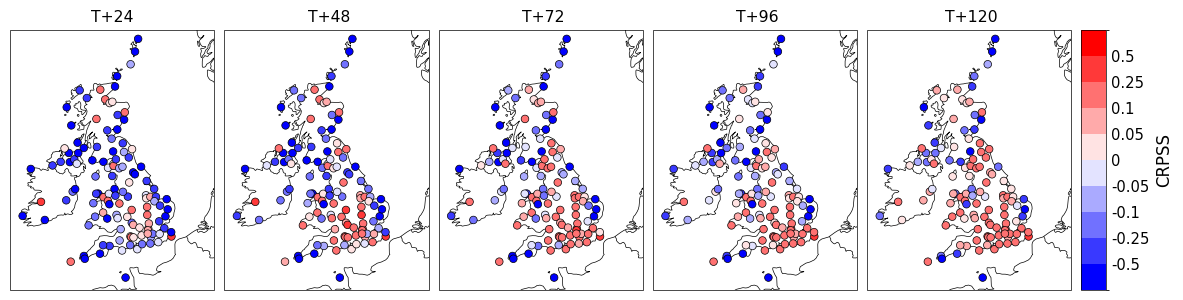}}
 \includegraphics[width=39pc]{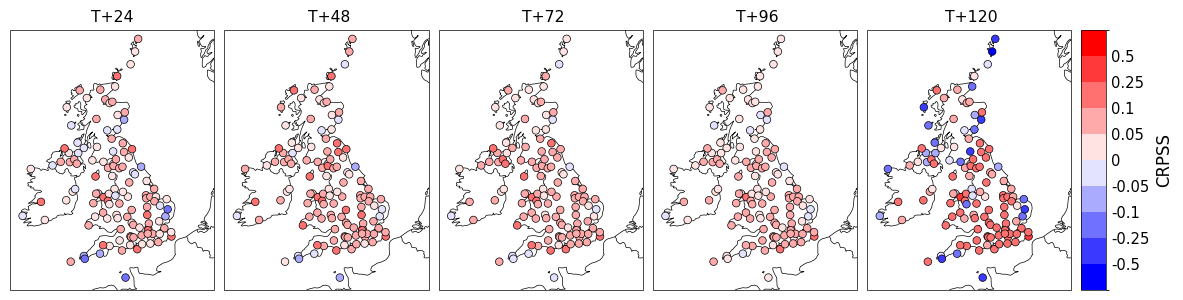}
	\caption{Continuous Ranked Probability Skill Score (CRPSS) at different lead times computed over all days with MOGREPS-UK being the reference model. Top row is for AIFS-CRPS and lower row is for the blend. Negative value (blue) means that the reference model is better and red means that it is worse.}\label{fig4} 
\end{figure}

So far, the overall performance of the blend has been assessed by averaging over all days and sites. This averaging hides spatial variations in the performance of the blend. Figure \ref{fig4} shows the Continuous Ranked Probability Skill Score computed over all days and using MOGREPS-UK as the reference model. Positive skill means that the ensemble being tested is better than MOGREPS-UK and negative skill means the MOGREPS-UK baseline is better. The top row shows that MOGREPS-UK outperforms AIFS-CRPS in most UK locations at shorter lead times for 2m temperature, though this skill differential decreases with lead time, and by T+120 AIFS-CRPS outperforms MOGREPS-UK in many locations.

However, the second panel demonstrates that by blending MOGREPS-UK with AIFS-CRPS, the blend outperforms MOGREPS-UK almost everywhere in the UK on day one, despite MOGREPS-UK's superior skill over AIFS-CRPS at this lead time. At most sites the blend gains more skill as the lead time increases, especially in central England and northern Scotland. Since the performance of individual models varies over locations, the implication might be that the blend can be further improved by optimizing weights with respect to lead times and locations, as done in \citet{chakraborty2025mowe}.

\subsection{Is the improvement in the forecast accuracy of the blend because it is an AI weather model?}
To better understand the value of adding an AI-based ensemble model into the blend, we consider a blend of two physics-based models alongside physics-based and AI-based combinations. MOGREPS-UK is blended with IFS, with weight $80\%$ assigned to the more skillful model at each lead time. For example, MOGREPS-UK is assigned a weight of $80\%$ from day one to day four and then $20\%$ at day five in the MOGREPS-UK and IFS blend. 
\begin{figure}[h]
\centerline{
 \includegraphics[width=33pc]{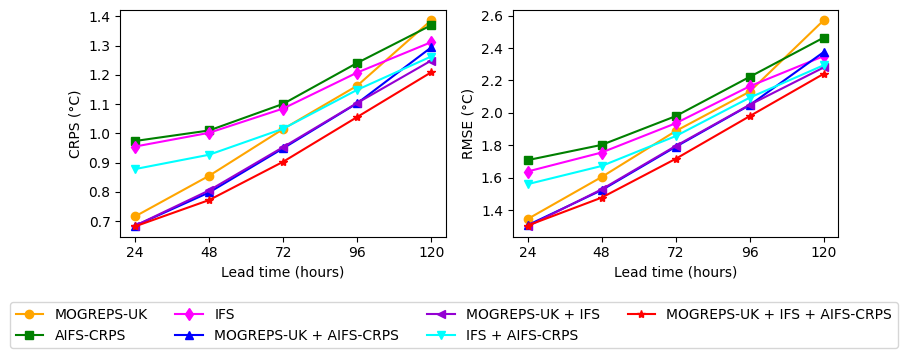}}
	\caption{CRPS and RMSE by lead time for individual models and their blends averaged over all days and sites. }\label{fig5} 
\end{figure}

Figure \ref{fig5} shows that IFS outperforms AIFS-CRPS at all lead times, but still underperforms MOGREPS-UK, except at day five. This figure also shows that there is a slight difference in overall skill between the blend of MOGREPS-UK and AIFS-CRPS and that of MOGREPS-UK and IFS. However, the blend of IFS and AIFS-CRPS underperforms the other blends across almost all lead times. This suggests that the improvement in the blend is not necessarily because AIFS-CRPS is an AI-based model, but because of the difference in model error characteristics, that is, model diversity matters. IFS and AIFS-CRPS are different models made by the same organization (ECMWF). AIFS-CRPS was trained on ERA5 data which was created from the IFS model, as such it is perhaps unsurprising that the models share some behavior, even though they are using completely different techniques for prediction. MOGREPS-UK is developed by the Met Office from the physics-based Unified Model \citep{walters2019met} without relationship to IFS, and this independence may add strength to the overall diversity of the blend.

The blend of all three models with the following weighting: $50\%$ MOGREPS-UK, $30\%$ IFS and $20\%$ AIFS, chosen based on forecast skill, gives the highest improvement, especially from day two to day five, as shown in Figure \ref{fig5}. This suggests that the more models there are in the blend, the better the overall forecast skill when appropriately weighted, as discussed in \citet{trotta2025statistical}.

\begin{figure}[h]
\centerline{
 \includegraphics[width=33pc]{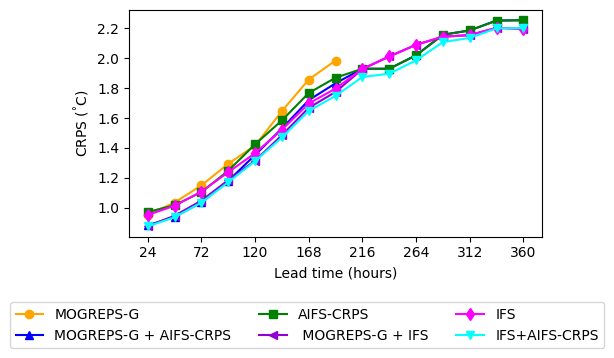}}
	\caption{CRPS by lead time for individual models and their blends averaged over all days and sites.}\label{fig6} 
\end{figure}

To explore the performance of the blends at longer lead times, we use MOGREPS-G, which is valid up to T+198. It is a lower resolution model than MOGREPS-UK and as such, does not perform as well for the UK domain. In this case, MOGREPS-G is less skillful than IFS and AIFS-CRPS over this domain. Despite MOGREPS-G performing less well than AIFS-CRPS and IFS, figure \ref{fig6} shows that blending all three models also outperform the best performing individual model at lead times up to T+198. Beyond this lead time, MOGREPS-G is unavailable. The only blend from T+198 to T+360 is that of IFS and AIFS, but this blend also outperforms individual models. 

We further investigate the differences in the blends by exploring their spread skill ratios \citep{whitaker1998relationship} over lead times. This ratio compares the ensemble spread to the ensemble skill. Here, we compute the mean ensemble variance divided by the mean squared error of the ensemble means, and then take the square root of that ratio to obtain the SSR. If SSR $=1$, then the forecast is well calibrated, while it is under-dispersed when SSR $< 1$ and over-dispersed when SSR $> 1$. For blended forecasts, we interpolate between the probabilities specified using multiple thresholds to generate percentiles that we use as ensemble members to compute the SSR \citep{flowerdew2014calibrating}.
\begin{figure}[h]
\centerline{
 \includegraphics[width=33pc]{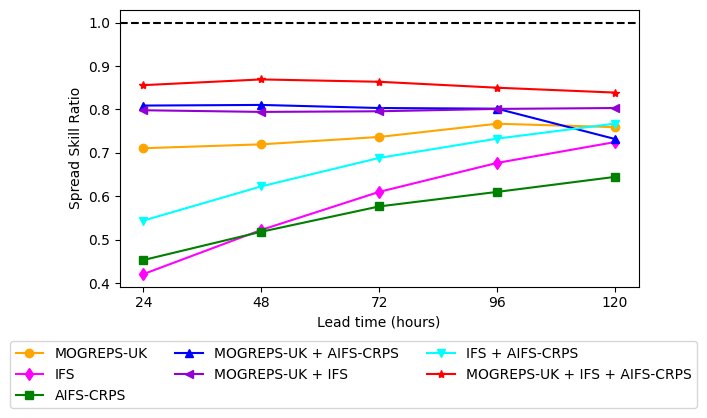}}
	\caption{Spread skill ratio (SSR) against lead time for individual models and their blends averaged over all sites.}\label{fig7} 
\end{figure}

Figure \ref{fig7} shows that MOGREPS-UK has the best SSR when comparing the individual models. For comparison of two-model blends, a blend of AIFS-CRPS and MOGREPS-UK shows the best SSR at all lead times, except for day five. This implies that blending AIFS-CRPS and MOGREPS-UK is slightly better at reducing under-dispersion than blending IFS and MOGREPS-UK. The drop in SSR in MOGREPS-UK on day five suggests that using only the value of CRPS to assign weights may not be the ideal methodology for creating an optimal blend. The three-model blend has the best SSR across all lead times, demonstrating the benefits of adding more appropriately weighted models into the blend.

\subsection{Impact of calibration on blending}
\begin{figure}[h]
\centerline{
 \includegraphics[width=33pc]{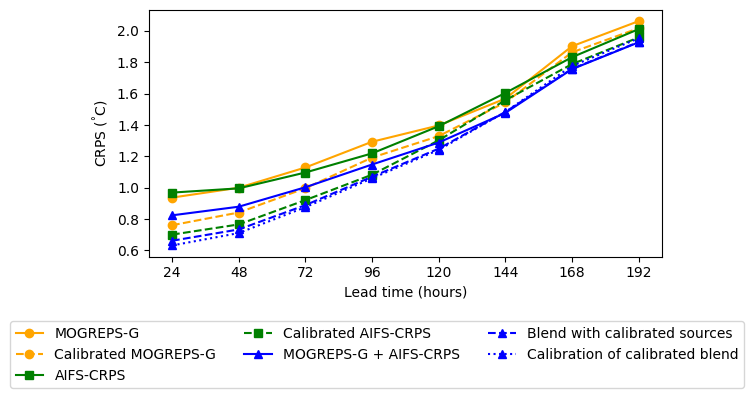}}
	\caption{CRPS by lead time for individual models and the blend of both averaged over all days and sites.}\label{fig61} 
\end{figure}

As discussed in the Introduction, calibration is used to improve the biases in weather forecast ensembles, leading to better forecast accuracy at surface sites. The impact of calibration before blending is investigated by applying EMOS to MOGREPS-G and AIFS-CRPS at all sites and lead times, as shown in Figure \ref{fig61}. The figure shows that calibration improves the skill of both ensemble physics-based and AI-based forecasts at all lead times. The result for MOGREPS-G forecasts is consistent with the findings of \citet{wessel2024lead}. \citet{kocsis2026ai} also found that calibrating AIFS-CRPS wind speed forecasts leads to increased skill, especially at short lead times. Our results therefore complement those of \citet{trotta2025statistical} for deterministic AIFS. Calibrating AIFS-CRPS forecasts leads to improvements similar to those of ensemble physics-based forecasts. Future research could compare the implementation of different calibration methods on both AI-based and physics-based forecasts in the spirit of \citet{kocsis2026ai}. 
Figure \ref{fig61} also shows that calibrating individual forecasts before blending improves the overall performance of the blend compared to the blend of the raw forecasts. Note that for this experiment a 50/50 weighting configuration is used in the blend, since the aim is to demonstrate the impact of calibration and not to obtain an optimal blend. 

\citet{ranjan2010combining} showed that blending calibrated forecasts does not necessarily give rise to a calibrated blended forecast. Figure \ref{fig61} shows that the application of EMOS again to the calibrated blend also improves forecast skill, especially at shorter lead times. An alternative method that can be used to calibrate blended probabilities is the so called beta recalibration approach proposed by \citet{gneiting2013combining}. The implication of the work shown here is that there is a benefit in applying calibration methods before and after blending, as discussed in \citet{trotta2025statistical}.
\subsection{Where does the improvement in the blend come from?}
To understand the source of improvement in the blend, we decompose the mean CRPS of the blend into components that describe the individual and cross model terms. A similar decomposition was proposed for the mean squared error by \citet{knuppel2022forecast} and \citet{kang2022forecast} for deterministic blends. For the special case of equal weights, \citet{engle1983estimates} and \citet{lahiri2010measuring} proposed the same decomposition for the mean squared error.  

The probabilistic error at a threshold $t$ for a forecast with CDF $F$ and its corresponding observation $y$ is given as
\begin{displaymath}
    e(t)=F(t)-\mathbbm{1}\{ t \geq y\}.
\end{displaymath}
The signed bias at the threshold $t$ of this error can be assessed by looking at its mean value. This bias measures the marginal calibration of the forecast. If the mean value is positive, it means that the forecast puts too much probability below $t$, and a negative value means that the forecast puts too little probability below $t$. The CRPS can be defined as an integral over all thresholds of squared errors as follows
\begin{displaymath}
    \text{CRPS}(F,y)=\int_{-\infty}^{+\infty}e^2(t)dt.
\end{displaymath}
Now, suppose that $F_1,\ldots,F_n$ are cumulative distribution functions for $n$ probability forecasts for a single observation $y$. Then a blended forecast by the linear pool \citep{stone1961pool} is defined as
\begin{displaymath}
    F(t)=\sum_{i=1}^{n}w_iF_i(t)
\end{displaymath}
where $\sum w_i=1$ for $w_i\geq 0$ and $i$ is the model index and $n$ is the total number of models in the blend. For a given model $i$, the error at the threshold $t$ is $e_i(t)=F_i(t)-\mathbbm{1}\{ t \geq y\}$ and the blend error is $e(t)=\sum_{i=1}^{n}w_ie_i(t)$. The square of this error can be written as
\begin{displaymath}
    e^2(t)=\left(\sum_{i=1}^{n}w_ie_i(t)\right)^2.
\end{displaymath}
But we know that the weighted error difference squared can be expanded as follows    
\begin{displaymath}
    \sum_{i=1}^{n}\sum_{j=1}^{n}w_iw_j[e_i(t)-e_j(t)]^2=2\sum_{i=1}^{n}w_ie_i^2(t)-2\left(\sum_{i=1}^{n}w_ie_i(t)\right)^2.
\end{displaymath}
Integrating the squared error over all thresholds yields the CRPS of the blend given as
\begin{equation}
\label{CRPS_blend}
    \mathrm{CRPS}(F,y)=\sum_{i=1}^{n}w_i\mathrm{CRPS}_i-\sum_{i<j}w_iw_jC_{i,j}
\end{equation}
where $\mathrm{CRPS}_i$ is the CRPS of the individual model and $C_{i,j}\geq 0$ is the cross term defined as 
\begin{displaymath}
    \mathrm{C}_{i,j}=\int_{-\infty}^{+\infty} [e_i(t)-e_j(t)]^2\,dt.
\end{displaymath}
The first term in Equation \eqref{CRPS_blend} is the weighted mean of the component scores, and the second term measures the diversity in the component distributional error. The cross term is equal to the divergence associated to the CRPS and measures the difference between the corresponding CDFs when evaluated at all thresholds \citep{thorarinsdottir2013using}. This term is also known as forecast disagreement in economic forecasting \citep{knuppel2022forecast,kruger2024kernel}. Equation \eqref{CRPS_blend} shows that the multi-model blend can achieve a lower (better) score than the weighted average of its components whenever the component errors (forecasts) differ sufficiently. The decomposition given in Equation \eqref{CRPS_blend} has been generalized to the class of kernel scores  \citep{gneiting2007strictly}, as shown in Proposition 5.1 of \citet{kruger2024kernel}. To understand the dissimilarity between forecast distributions, \citet{resin2024shift} decomposed the cross term into shift and dispersion components. The shift components quantify differences in location, while the dispersion components measure differences in variability between forecasts $F_i$ and $F_j$. Their decomposition of the Cram\'{e}r distance in Proposition 2.7 is identical to the decomposition of our cross term, and this sort of analysis can be explored to understand how forecasts differ in the diversity term. 

If two models have identical errors, then the cross term is zero, and blending leads to little or no benefits. If the errors are complementary, that is, large errors of opposite signs, then the diversity term becomes large, and blending helps to improve the overall forecast performance. The accuracy of the blend can be improved by weighting the model with the smallest CRPS more and that with the largest CRPS less, as this will reduce the CRPS of the individual terms in the CRPS of the blend. Another strategy to improve accuracy is to dynamically weight the models to minimize CRPS at different thresholds and lead times. 
\begin{figure}[h]
\centerline{
 \includegraphics[width=39pc]{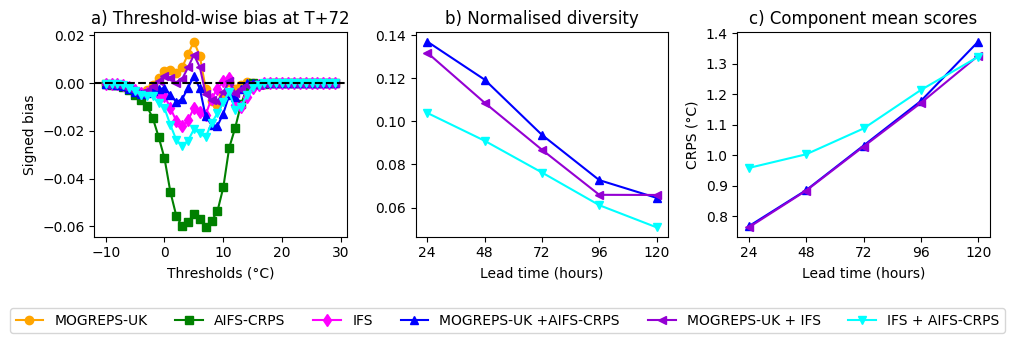}}
	\caption{(a) Signed probability bias against thresholds at T+72 lead time where a score of zero is ideal (b) Normalized diversity against lead time where a higher score is more diserable and (c) Component mean scores against lead time for individual models and their blends averaged over days and sites, where a lower score is better.}\label{fig611} 
\end{figure}

Figure \ref{fig611}a shows the mean signed bias at 72 hour lead time at different temperature thresholds averaged over all sites for MOGREPS-UK, AIFS-CRPS, IFS and their corresponding blends. The figure shows that MOGREPS-UK is positive for lower thresholds and negative for higher thresholds, suggesting marginal over-dispersion. The blend of MOGREPS-UK and IFS shows a pattern similar to that of MOGREPS-UK. The others are mostly negatively biased, which implies that their forecasts tend to have higher temperatures than the observations. 

Figure \ref{fig611}b shows a plot of the mean diversity divided by the mean CRPS of the blend at different lead times. This normalized diversity term helps us to quantify the contribution of the diversity term in the blend, that is, the proportion by which the CRPS of the blend is less than the mean of the component scores. The figure shows that the diversity contribution is highest in the MOGREPS-UK and AIFS-CRPS blend followed by MOGREPS-UK and IFS, and it is lowest in the blend of IFS and AIFS-CRPS. For example, at T+48, diversity reduces the overall CRPS in the MOGREPS-UK and AIFS-CRPS blend by approximately 12$\%$, while the reduction in the MOGREPS-UK and IFS blend is approximately $11\%$. However, at this lead time, the reduction in the blend of IFS and AIFS-CRPS is about $9\%$. Although there is a difference in the normalized diversity of the blend of MOGREPS-UK with AIFS-CRPS and MOGREPS-UK with IFS, this difference has only a very minor impact on the overall skill, as shown in Figure \ref{fig5}. This is because the magnitude of this difference is small and the value of the overall skill is dominated by the component mean scores, as shown in Figure \ref{fig611}c.

Figure \ref{fig611}c shows the weighted component means of the different blends at different lead times. This figure shows that the accuracy of the MOGREPS-UK blend with AIFS-CRPS or IFS mostly outperforms that of the IFS and AIFS-CRPS blend. This is because MOGREPS-UK is the most skillful forecast and has been weighted more in these blends. Also, blending MOGREPS-UK with AIFS-CRPS or IFS does not produce much difference in their component means, since they have similar skills, especially at shorter lead times. In general, apart from assigning weights based on the model performance, optimal weights can be derived by minimizing the CRPS subject to the constraint on the weights using the so called Lagrange multiplier formalism \citep{timmermann2006forecast,newbold1974experience}. Because linear pooling naturally inflates variance (dispersion errors), non-linear methods, such as the Beta transform linear pool, are often used to scale the combined CDFs to preserve calibration \citep{gneiting2013combining}.

\section{Discussion and Conclusions}
Weather forecast organizations worldwide are interested in creating the most accurate forecast for their chosen user domain. Using a multi-model blend of physics-based ensemble models has long been accepted as the best strategy to achieve accuracy, especially if high resolution models are unavailable. The work presented here sets out to understand what the impact on accuracy is of adding an AI model ensemble with a physics-based ensemble.

\citet{trotta2025statistical} used fuzzy thresholding to construct probabilistic forecasts from deterministic AIFS and showed that it gives improvements in skill when combined with physics-based models. Using the Met Office's IMPROVER post-processing system, we have shown that a blend of AIFS-CRPS and MOGREPS-UK yields an increase in forecast skill in terms of both the deterministic and probabilistic forecasts whenever the weighting is based on relative forecast skill. In addition, a blend of all three models, MOGREPS-UK, IFS, and AIFS-CRPS, yields the highest increase in overall forecast skill across all lead times when weighted according to relative forecast skill.  The increase in overall temperature forecast skills in the blend is also observed after week one and for wind speed forecasts. 

A Murphy decomposition of the mean Brier score at chosen thresholds reveals that blending improves both the reliability and resolution of the forecast. This decomposition can also help identify which property of the forecast needs to be improved, through post-processing, before and after blending.

Blending forecasts from models with similar error characteristics yields a smaller increase in overall forecast skill compared to blending forecasts with different characteristics. For example, a blend of MOGREPS-UK and AIFS-CRPS or IFS is more skillful than a blend of IFS and AIFS-CRPS which have similar characteristics. Our findings show that a blend of MOGREPS-UK and AIFS-CRPS has a better spread skill ratio than a blend of two physics-based models, MOGREPS-UK and IFS. This implies that pairing an AI-based model with a physics-based model with more individual skill but also with more complementary errors leads to a better overall performance. A decomposition of the CRPS of the blend into individual and cross terms shows that models with complementary errors have larger diversity terms, hence smaller CRPS of the blend.

This has interesting implications for model development and highlights the benefit of model diversity. The findings demonstrate that different behavior in errors strengthens the effectiveness of the multi-model forecast. This aligns with previous work on multi-model ensembles (e.g \citet{krishnamurti2016review, neal2024seamless}), but we believe this is the first time it has been demonstrated by blending AI and physics-based models and also points to the potential advantage of having models that are not based on IFS data (and by connection, ERA5) in the blend. 

The spatiotemporal variation of the performance of the blend calls for the use of dynamic weighting in both time and space. \citet{trotta2025statistical} used optimal weights for each forecast source deduced by optimizing the combination of forecasts available at each lead time. Therefore, future studies will explore the use of optimal weights that minimize a metric such as MSE or CRPS at each lead time and location. The decomposed CRPS shows that a more accurate blend can be achieved by weighting models with a smaller CRPS more and those with a larger CRPS less.

The application of EMOS on AIFS-CRPS temperature forecasts leads to a similar increase in forecast skills, as observed in \citet{kocsis2026ai} for wind speed forecasts. Also, calibrating the forecast sources before and after blending increases the overall skill of the blended forecast. Future research could explore the impact of different calibration methods on forecast sources and blended forecasts. 

%

%

\acknowledgments
This work was funded by the Met Office Supercomputer Grant 20393746 for the improvement of weather forecasts for the UK public good.
%
%
\availstatement
Data for raw AIFS-CRPS and IFS forecasts are available from ECMWF via the MARS API. The open-source IMPROVER postprocessing software is available from https://github.com/metoppv/improver/.








%



\appendix
\section{The CRPS and RMSE of wind speed for the evaluation period (November 2024 to January 2025)}    
\begin{figure}[h]
\center
 \includegraphics[width=33pc]{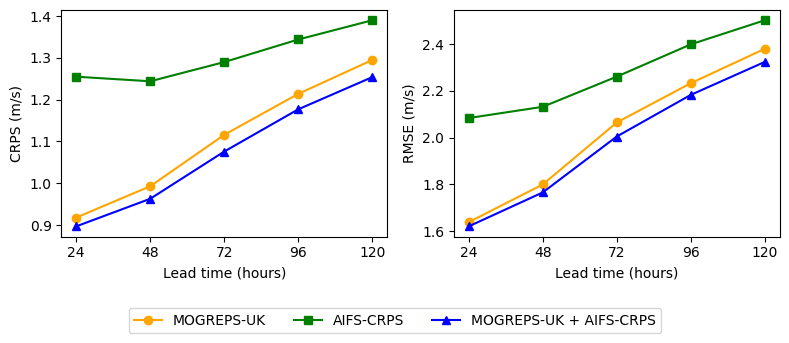}
	\caption{CRPS and RMSE by lead time for individual models and the blend of both averaged over all days and sites. }\label{appen1} 
\end{figure}

\clearpage
\section{The CRPS and RMSE of temperature for the month of July 2025}
\begin{figure}[h]
\center
 \includegraphics[width=33pc]{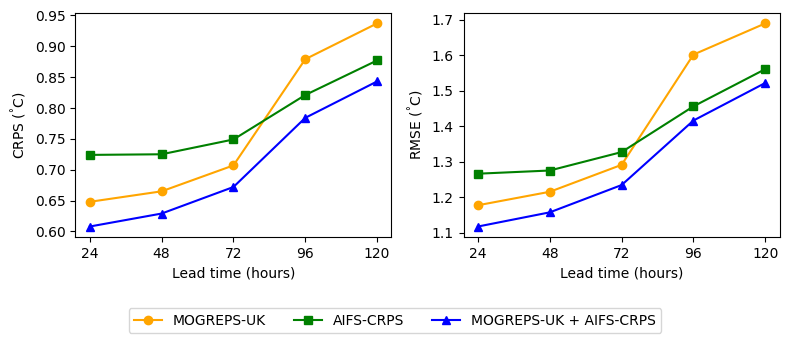}
	\caption{CRPS and RMSE by lead time for individual models and the blend of both averaged over all days and sites.}\label{appen2} 
\end{figure}

\clearpage

\bibliographystyle{ametsocV6}
\bibliography{references}

\end{document}